# Controllable CVD-Growth of 2D $Cr_5Te_8$ Nanosheets with Thickness-Dependent Magnetic Domains

*Hanxiang Wu[1‡], Jianfeng Guo[1‡], Suonan Zhaxi[1], Hua Xu[1], Shuo Mi[1], Le Wang[1], Shanshan Chen[1], Rui Xu[1], Wei Ji[1], Fei Pang[1]\* and Zhihai Cheng[1]\**

[1]Beijing Key Laboratory of Optoelectronic Functional Materials & Micro-nano Devices, Department of Physics, Renmin University of China, Beijing 100872, China

**ABSTRACT:** As a typical 2D magnetic material with self-intercalated structure, $Cr_5Te_8$ exhibits many fascinating magnetic properties. Although the ferromagnetism of 2D $Cr_5Te_8$ has been reported, the study of its magnetic domain is still blank. Herein, we successfully fabricated 2D $Cr_5Te_8$ nanosheets with controlled thickness and lateral size by chemical vapor deposition (CVD). Then magnetic property measurement system suggested $Cr_5Te_8$ nanosheets possessing intense out-of-plane ferromagnetism with the Curie temperature of 179 K. Most importantly, we found magnetic bubbles and thickness-dependent maze-like magnetic domains by cryogenic magnetic force microscopy (MFM) for the first time. The domain width of the maze-like magnetic domains increases rapidly with decreasing sample thickness, while domain contrast decreases. This indicates dipolar interaction is the dominant role over exchange interaction and magnetic anisotropy. Our work not only paves a way for the controllable growth of 2D magnetic materials, but also suggests new directions for controlling magnetic phases and systematically adjusting domain properties.

‡ These authors contributed equally to this work.

\* To whom correspondence should be addressed: feipang@ruc.edu.cn, zhihaicheng@ruc.edu.cn

# 1. INTRODUCTION

As an essential part of the 2D material family, due to the great application prospect in spintronics and magnetic memory devices,[1,2] 2D magnetic materials with excellent electrical[3,4] and photoelectrical[5,6] properties as well as novel magnetic[7-9] properties have received extensive attention. Particularly, it is found that monolayer $Fe_3GeTe_2$ still possessed ferromagnetic long-range order at low temperature.[10-12] To date, ultrathin $Cr_xGa_{1-x}Te$ nanosheets exhibited highly tunable, intrinsic room-temperature ferromagnetism and indicated their preferable potential for carrier control.[13] Although 2D magnetic materials hold fascinating prospects for exploration and practical applications, its magnetic behavior especially in the magnetic domain structure is still in its infancy. Therefore, it is very important to explore the controllable growth and the magnetic domain structure of 2D magnetic materials.

Notably, Cr-based chalcogenides have a wide self-intercalated phases and thickness-dependent properties due to their abundant compositions and structures and novel magnetic properties for fundamental study and promising technological applications.[14-31] Particularly, 2D $Cr_5Te_8$ nanosheet was demonstrated to be ferrimagnetic with strong out-of-plane spin polarization, and the observed Curie temperature increases monotonously from 100 K in the thin flake (10 nm) to 160 K in the thick flake (30 nm), which results from the critical role of the strong interlayer coupling of the magnetic order,.[32-38] Moreover, the controllable growth of 2D $Cr_5Te_8$ with high quality is still an immense challenge. Chemical vapor deposition (CVD) method has unique advantages in the preparation and phase-selective growth of non-layered materials. For example, multifarious kinds of non-layered nanoflakes such as $Cr_2S_3$,[39-43] $MnSe$,[44,45] $Fe_7Se_8$[46] and $CoSe$[47] have been obtained by CVD. Hence, CVD may provide a possible route to prepare nonlayered 2D $Cr_5Te_8$ nanosheets with high quality.

Herein, we reported the controllable growth of 2D self-intercalated $Cr_5Te_8$ nanosheets via CVD and found temperature and source–substrate distance are key parameters for synthesis of 2D $Cr_5Te_8$ nanosheets. Furthermore, we found two different kinds of magnetic domain structures: magnetic bubbles and thickness-dependent maze-like magnetic domains by magnetic force microscopy (MFM). Optical microscopy (OM), atomic force microscopy (AFM), X-ray diffraction (XRD), scanning electron microscopy (SEM) with energy dispersive spectroscopy (EDS), Raman and X-ray photoelectron spectroscopy (XPS) measurements illustrated the high crystallinity and accurate composition of the grown 2D $Cr_5Te_8$ nanosheets. Furthermore, magnetic property measurement system (MPMS) measurements demonstrated that $Cr_5Te_8$ nanosheets possess both out-of-plane and in-plane ferromagnetism with the Curie temperature of 160 to 179 K, which was lower than those previously reported in bulk samples.[48] Moreover, we also studied the thickness-dependent maze-like magnetic domains by MFM. The first discovery of the strong thickness-dependent behavior of the domain width and domain contrast of the maze magnetic domain in 2D $Cr_5Te_8$ provides a good prospect for their practical application in magnetic memory devices and other ferromagnetism-pertinent fields. Meanwhile, this work would also provide a reference for the controllable growth and the research of the magnetic domain structure of other non-layered 2D magnetic materials and pave a way for further research on these 2D magnetic materials and their magnetism-pertinent applications in spintronics and magnetic memory devices.

## 2. RESULTS AND DISCUSSION

The schematic diagram of CVD growth process of $Cr_5Te_8$ nanosheets is described in Figure 1a (Figure S1, Supporting Information), which is used to control thickness, nucleation density and lateral size of $Cr_5Te_8$ nanosheets on mica substrates and $SiO_2$/Si substrates. The crystal structure of $Cr_5Te_8$ is shown in Figure 1b and c, which has a self-intercalated crystal structure intercalating Cr atoms into the van der Waals gap between the $CrTe_2$ layers.[37,38] Typical OM images of $Cr_5Te_8$ are shown in Figure 1d and e, respectively. The $Cr_5Te_8$ nanosheets are mainly triangular on mica and hexagonal on $SiO_2$/Si. The $Cr_5Te_8$ nanosheets on mica are universally thinner than those on $SiO_2$/Si, which is attributed to numerous suspended bonds on $SiO_2$/Si substrates. We also fabricated 2D $Cr_5Te_8$ on sapphire substrates and the as-grown $Cr_5Te_8$ nanosheets on different substrates (Figure S2, Supporting Information). The thickness of $Cr_5Te_8$ nanosheets as shown in Figure 1f and g. Compared with the c-axis length 0.60 nm, the grown sample in Figure 1f is a bilayer nanosheet with a lateral size of approximately 25 μm. OM and SEM images corresponding to Figure 1f and g are presented in Figure S3 (Supporting Information). AFM morphology and OM image of a typical hexagonal $Cr_5Te_8$ nanosheet with a thickness of 9.4 nm grown on the mica substrate are shown in Figure S4 (Supporting Information).

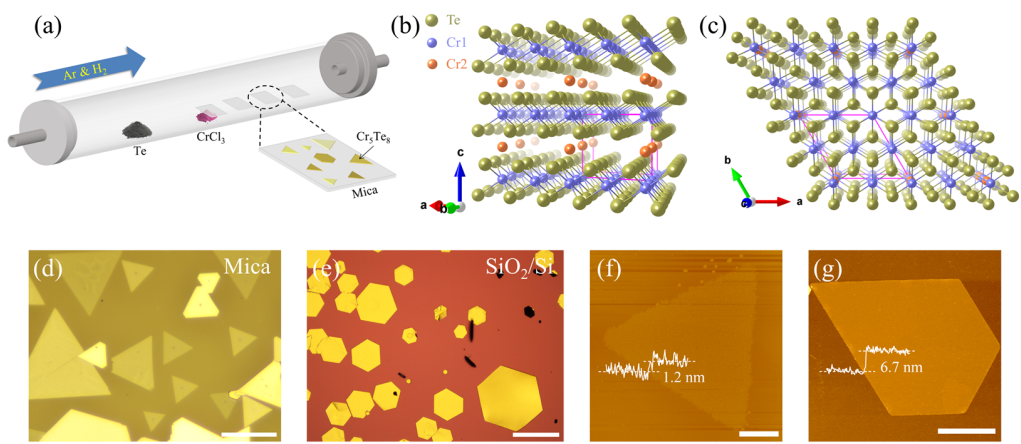

**Figure 1.** Growth method and morphology characterization of $Cr_5Te_8$ nanosheets. (a) Schematic diagram of the CVD setup for synthesis of $Cr_5Te_8$ nanosheets. (b,c) Crystal structure (top (b) and side (c) views) of $Cr_5Te_8$ obtained by CVD. (d,e) Typical OM images of $Cr_5Te_8$ grown on mica (d) and $SiO_2$/Si (e) substrates. The scale bars are 30 and 200 μm, respectively. (f,g) AFM images

and corresponding height profiles of Cr$_5$Te$_8$ nanosheets grown on mica (f) and SiO$_2$/Si (g) substrates. The scale bars are 5 μm.

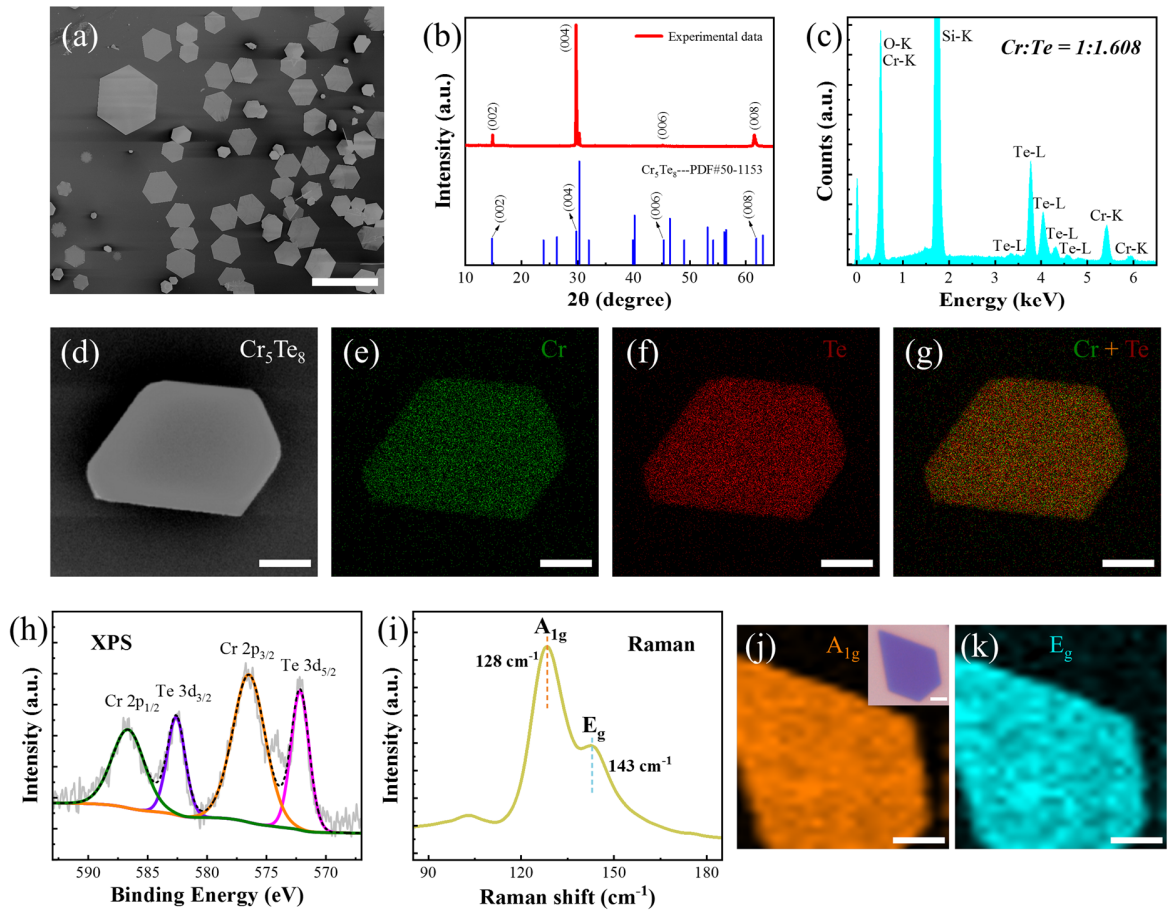

**Figure 2.** Composition and structure characterization of CVD-grown Cr$_5$Te$_8$ nanosheets. (a) Large-scale SEM image of Cr$_5$Te$_8$ nanosheets. The scale bar is 300 μm. (b) XRD patterns of as-grown Cr$_5$Te$_8$ nanosheets (top) and the standard Cr$_5$Te$_8$ crystal from PDF card (bottom). (c) EDS analysis of CVD-grown Cr$_5$Te$_8$ nanosheets. (d-g) SEM image (d) of a hexagonal Cr$_5$Te$_8$ nanosheet and corresponding EDS mapping of Cr element (e), Te element (f) and an overlay of Cr and Te elements (g). The scale bars are 2 μm. (h) XPS of as-grown Cr$_5$Te$_8$ nanosheets. (i) Raman spectra of a Cr$_5$Te$_8$ nanosheet with a thickness of 6.7 nm. (j,k) Raman mapping

correlated with Raman peaks at 128 cm$^{-1}$ (j) and 143 cm$^{-1}$ (k) corresponding to $A_{1g}$ and $E_g$ mode. The inset shows the corresponding OM image of the Cr$_5$Te$_8$ nanosheet. The scale bars are 2 μm.

To examine the crystal phase of the as-synthesized Cr$_5$Te$_8$ nanosheets, the composition and structure were characterized. A large-scale SEM image of Cr$_5$Te$_8$ nanosheets with a maximum lateral size up to 300 μm is shown in Figure 2a. To examine the crystal phase of the as-synthesized Cr$_5$Te$_8$ nanosheets, XRD characterization was performed. XRD pattern of 2D Cr$_5$Te$_8$ is shown in Figure 2b. Three diffraction peaks located at 15.14°, 30.12°, and 62.04° are indexed to the (002), (004), and (008) diffraction planes which matches well with the trigonal Cr$_5$Te$_8$ standard PDF card (PDF#50-1153). It illustrates the accuracy of the structure of as-grown 2D Cr$_5$Te$_8$.[37,38] Furthermore, only (00X) peaks appear in the XRD pattern, which suggested the surface of the Cr$_5$Te$_8$ crystal is parallel to the ab-plane. Further EDS was employed to analyze the elemental composition of the CVD-grown Cr$_5$Te$_8$ nanosheets. As shown in Figure 2c, the atomic ratio of Cr to Te is approximately 1: 1.61, which is extremely close to 5: 8, consistent with the stoichiometric ratio of Cr$_5$Te$_8$. EDS mapping of Cr and Te elements is shown in Figure 2d, e, f and g, whose uniform color distribution manifests the formation of Cr$_5$Te$_8$ crystal and the uniformity of spatial distribution of its components. XPS was used to examine the chemical composition and bonding type of the CVD-grown Cr$_5$Te$_8$. As shown in Figure 2h, the peaks located at binding energies of ≈576.47 and 586.64 eV are attributed to Cr 2p$_{3/2}$ and Cr 2p$_{1/2}$, and the peaks located at ≈572.17 and 582.6 eV are attributed to Te 3d$_{5/2}$ and Te 3d$_{3/2}$, indicating a Cr$^{3.2+}$ state and Te$^{2-}$ state,[32] consistent well with the Cr$_5$Te$_8$ crystal, respectively. Eventually, Raman spectra was performed in a Cr$_5$Te$_8$ nanosheet with a thickness of 6.7 nm, as presented in Figure 2i. Two prominent peak positions at 128 cm$^{-1}$ and 143 cm$^{-1}$ corresponding to $A_{1g}$ and $E_g$ mode appear for the Cr$_5$Te$_8$ nanosheet, separately. Raman mapping correlated with Raman peaks at 128 cm$^{-1}$ and 143 cm$^{-1}$ corresponding to A$_{1g}$ and E$_g$ mode is shown in Figure 2j and k, whose homogeneous contrast with background further demonstrates the high uniformity of as-grown Cr$_5$Te$_8$ nanosheets. Briefly, all these results confirm that 2D Cr$_5$Te$_8$ nanosheets with high-quality and accurate composition were successfully synthesized, which ensures a brilliant material foundation for further research on the controllable growth and magnetic behavior.

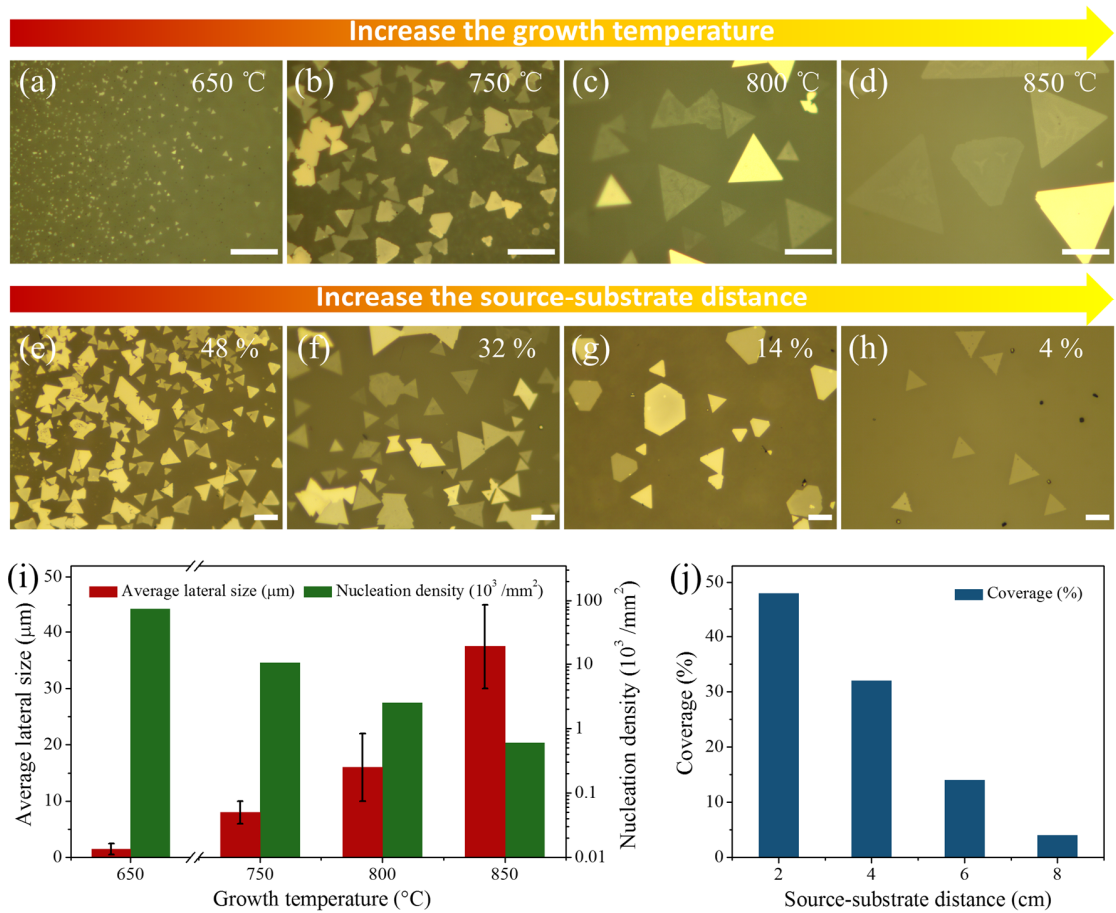

**Figure 3.** Controllable growth of Cr$_5$Te$_8$ nanosheets on mica. (a-d) OM images of Cr$_5$Te$_8$ nanosheets at different growth temperature 650 °C (a), 750 °C (b), 800 °C (c) and 850 °C (d), respectively. (e-h) OM images of Cr$_5$Te$_8$ nanosheets synthesized under the same conditions (grown at 800 °C) with coverage of ≈ 48 %, 32 %, 14 % and 4 % for different source–substrate distance ($D_{ss}$) of ≈ 2 cm (e), 4 cm (f), 6 cm (g) and 8 cm (h), respectively. Scale bars: 20 μm. (i)

Histogram of average lateral size (red bar) and nucleation density (green bar) with variational growth temperature corresponding to (a-d). (j) Histogram of coverage (blue bar) with variational source–substrate distance corresponding to (e-h).

In order to achieve controllable growth of $Cr_5Te_8$ nanosheets, systematic experiments were carried out. It's found growth temperature and source–substrate distance ($D_{ss}$) between the $CrCl_3$ source and mica substrate are crucial to control average lateral size, nucleation density and coverage of $Cr_5Te_8$ nanosheets. To investigate the effect of growth temperature on the growth results, the other growth parameters (carrier gas and $D_{ss}$=4 cm) remains constant and the temperature of Te powder remains constant at 550 °C. $Cr_5Te_8$ nanosheets on mica grown at different temperature exhibit an obvious change in average lateral sizes and nucleation densities. Typical OM images of $Cr_5Te_8$ grown on mica at different growth temperature are presented in Figure 3a–d. When the growth temperature is 650 °C, the especially small average lateral size about 1.5 μm and high nucleation density of about 72,500/mm$^2$ remains. For the temperature (650 °C) is lower than the evaporation temperature of $CrCl_3$ source, there is no sufficient amount of the reactants. Increasing growth temperature to 750 °C, the average lateral size remains rapidly increasing up to about 8 μm, and the nucleation density dramatically reduces to 10,500/mm$^2$, which implies an adequate supply of reaction sources at the suitable growth temperature. At higher growth temperature (800 °C), the average lateral size continues to increase up to about 16 μm, and the nucleation density is continuously reduced at 2,500/mm$^2$. When the growth temperature is kept considerably high at 850 °C, the average lateral size gets extremely big, approximately 37.5 μm, and the comparatively exceptionally low nucleation density of about 600/mm$^2$ is obtained, which provides a valuable reference for the growth of single crystals with large transverse size. As shown in Figure 3i, growth temperature has a significant effect on the $Cr_5Te_8$ size and nucleation density. These growth regulation rules accord with the nucleation model of vapor deposition presented by W. K. Burton *et. al.*, which shows that growth temperature is positively correlated with transverse size, but negatively with nucleation probability.[49] At lower growth temperature, the growth is primarily controlled by kinetics, resulting in a higher nucleation density and smaller nanosheets. At higher growth temperature, the growth is mainly controlled by thermodynamics, which reduces the nucleation density to generate larger nanosheets.

Furthermore, we found the distance between the $CrCl_3$ source and the mica substrates ($D_{ss}$) was also an essential parameter to control the coverage of $Cr_5Te_8$ nanosheets. With the same growth parameters (carrier gas and growth temperature) remaining constant and the growth temperature remaining constant at 800 °C, the $Cr_5Te_8$ nanosheets on mica synthesized at different $D_{ss}$ exhibit a distinct change in their coverages shown in Figure 3e–h. As shown in the histogram of Figure 3j, with increasing $D_{ss}$ from 2 cm to 8 cm, the coverage of $Cr_5Te_8$ on the mica substrates decreased from 48% to 4%. It is attributed to the low effective source concentration with increasing distance. Thus, the coverage between upstream and downstream locations is different. The remarkable effect of source–substrate distance on the growth process of $Cr_5Te_8$ has also been observed in other 2D materials.[18]

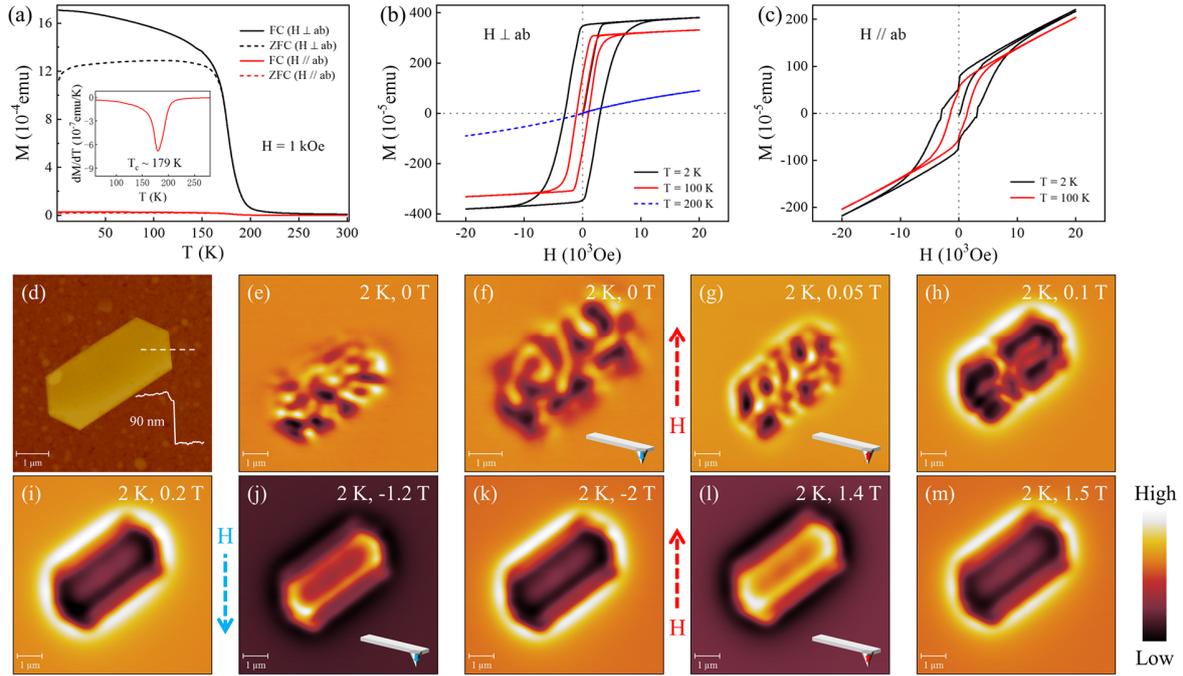

**Figure 4.** Ferromagnetism of $Cr_5Te_8$ nanosheets. (a) Temperature-dependent magnetic moments $M$–$T$ with the magnetic field perpendicular (black line) and parallel (red line) to the ab-plane for both FC (solid line) process and ZFC (dashed line) process on the sample under an external field of 1000 Oe. The insets show the first-order derivatives of $M$ with respect to $T$ to derive the $T_c$ of $Cr_5Te_8$. (b,c) Magnetic-field-dependent magnetic moments $M$–$H$ at temperatures of 2 K (solid black line), 100 K (solid red line) and 200 K (dashed blue line) with the magnetic field perpendicular (b) and parallel (c) to the ab-plane on the sample. (d) AFM topography of a $Cr_5Te_8$ single nanosheet. (e-m) Maze-like magnetic domains (e, f) and evolution (g-m) of the $Cr_5Te_8$ single nanosheet at the temperature of 2 K after a ZFC procedure with the external magnetic field ranged from -2 T to 1.5 T. The color scales are 23 (e, f), 32 (g, j), 52 (h, i), 74 (k, m) and 92 (l) deg.

In order to probe the magnetic properties of as-synthesized $Cr_5Te_8$ nanosheets, the magnetic susceptibility of 2D $Cr_5Te_8$ on mica were measured by MPMS. To avoid possible magnetic contamination, the samples were handled only with plastic tweezers. As shown in Figure 4a, the temperature-dependent magnetization measurements ($M$–$T$) with the magnetic field perpendicular ($H \perp ab$) and parallel ($H \,/\!/\, ab$) to the $ab$-plane for both field-cooling (FC) process and zero-field-cooling (ZFC) process were performed on the samples under an external field of 1 kOe. The $M-T$ data demonstrate a paramagnetic to ferromagnetic transition below the Curie temperature, $T_C$, which indicates the apparent ferromagnetic nature of synthesized 2D $Cr_5Te_8$ samples. The first-order derivative of $M$-$T$ in inset demonstrates that $Cr_5Te_8$ nanosheets possess a magnetic phase transition at $T_C \sim 179$ K, which are lower than those previously reported in bulk samples.[48] Additionally, the magnetization perpendicular to the $ab$-plane ($H \perp ab$) are much stronger than the parallel ($H \,/\!/\, ab$), indicating the $c$-axis is the magnetic easy axis and ferromagnetic order with enhanced perpendicular magnetic anisotropy.

There are several possible mechanisms for the origin of the reduced Curie temperature. Firstly, the increased proportion of surface atoms may lead to a slight change in the stoichiometry of Cr and Te elements, which can be regarded as a doping effect. The doping effect can tune ferromagnetism, and the thinner the nanosheet is, the higher the ratio of atoms on the surface accounts, the more noteworthy the tuning manifests resulting the more significant Curie temperature deviation from the bulk.[50,51] This is accordant with Haraldsen's pioneering theory, in which the Curie temperature of chromium telluride system shows an excessively gigantic sensitivity to Te element content.[50] Secondly, surface atoms may undergo reconstruction that changes the interatomic distance and thus the ferromagnetic interaction, as previously reported in two-dimensional $Cr_2Te_3$.[28] Thirdly, ferromagnetic interaction may change due to the stress imposed on $Cr_5Te_8$ nanosheets originated from the mismatch lattice constant between the mica substrate and $Cr_5Te_8$ nanosheets, leading to a change in Curie temperature.[27]

The magnetic field dependent magnetic moments (M–H) further confirmed the ferromagnetism and perpendicular magnetic anisotropy. M–H at different temperature with $H\perp ab$ and $H // ab$ are presented in Figure 4b and c, respectively. Large magnetic hysteresis loops when $H\perp ab$ and much smaller loops when $H // ab$, further confirmed that the c-axis is the magnetic easy axis. M–T and M–H of the thinner samples grown on a mica substrate and a SiO$_2$/Si substrate are shown in Figure S5 (Supporting Information), indicating a Curie temperature of 160 K and 169 K, respectively. It is consistent with the magnetic phase diagram of the layer number versus the temperature of Cr$_5$Te$_8$ single crystals.[32]

Since the thickness of Cr$_5$Te$_8$ nanosheets are not fully uniform, MPMS results are an averaged magnetic signal of the sample. In order to determine the ferromagnetic signal in a single Cr$_5$Te$_8$ nanosheet with well-defined thickness, the magnetic domain structures of Cr$_5$Te$_8$ nanosheets were characterized by cryogenic magnetic force microscopy (MFM). With nanoscale magnetic signal resolution, the MFM can provide real-space magnetic domain signals from individual Cr$_5$Te$_8$ nanosheets.[52]

The AFM topography of a 90 nm thick hexagonal Cr$_5$Te$_8$ nanosheet characterized by MFM is shown in Figure 4d. Figure 4e-m (Figure S6, Supporting Information) show the evolution process of its magnetic domain structure under different magnetic field. To observe the magnetic structures, the nanosheet was first cooled by ZFC from room temperature to 2 K. Obviously, the magnetic domain structure is different after every ZFC processes, are shown in Figure 4e (after the first ZFC procedure) and Figure 4f (after another ZFC procedure). This random formation process of magnetic domains confirms that the signals measured by MFM are indeed magnetic signals. In addition, the domain structure formed by spontaneous magnetization of the nanosheet after the ZFC process is a maze-like domain. Each domain of the $Cr_5Te_8$ nanosheet is about 3 μm in the length, 280 nm in the width, and 19 deg in the domain contrast.

After that, an external magnetic field perpendicular to the *ab* plane was applied to the $Cr_5Te_8$ nanosheet at 2K and the dynamic evolution of the maze-like magnetic domains was characterized. When the magnetic field increased gradually from 0 to 0.2 T, the magnetic domain changes greatly, as shown in Figure 4f-i. As the magnetic field increases, the area of the magnetic domains parallel to the direction of the field gradually expands. When the additional magnetic field is increased to 0.1 T, the dark contrast domains cover most of the nanosheet, with only a few bright contrast maze-like domains remaining in the central region of the nanosheet. When the additional magnetic field value increases to 0.2 T, the dark contrast domain region further expands and reaches saturation, forming a full dark single domain that fills the entire nanosheet. The contrast in this single domain is not uniform, with brighter contrast in the central regions and darker contrast in the edge regions, probably due to the anisotropy of the nanosheet shape and the broken translational symmetry in the edge regions.

Next, the magnetic field direction was reversed, and when the field was increased to -2 T, the MFM image also reversed, from full brightness to full darkness, indicating a reversal of the magnetization direction of the nanosheet. The direction of the additional magnetic field then reverses again. From 1.4 T to 1.5 T, the magnetic signal as a whole is reversed and the reversal of the magnetic domains does not evolve dynamically. The above evolution process of the magnetic domain of the $Cr_5Te_8$ single nanosheet with the applied magnetic field is completely consistent with the changing trend of the *M-H* curve of the mass magnetic signal of multi-nanosheets, indicating that our as-grown $Cr_5Te_8$ nanosheet has a distinguished hard magnetism.

In addition, Figure 4l and m also show that the coercivity field of the $Cr_5Te_8$ nanosheet is between 1.4 T and 1.5 T.

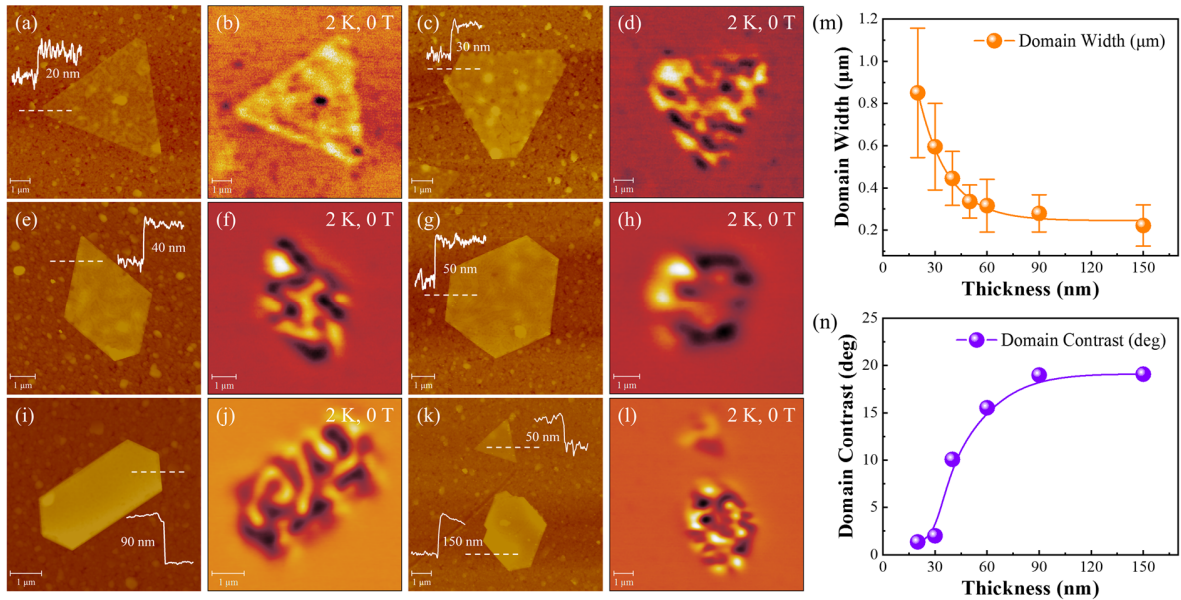

**Figure 5.** Maze-like magnetic domains of $Cr_5Te_8$ nanosheets with different thickness. (a-l) AFM topographies (a, c, e, g, i, k) and corresponding MFM images (b, d, f, h, j, l) of $Cr_5Te_8$ nanosheets with different thickness at 2 K. The color scales are 2.6 (b, d), 10 (f), 7.2 (h), 23 (j) and 19 (l) deg. (m,n) Domain width (m) and domain contrast (n) of the maze-like magnetic domains as variables of the thickness of $Cr_5Te_8$ nanosheets.

Then the magnetic domain structures of $Cr_5Te_8$ nanosheets with different thickness were systematically charactered by MFM, and a thickness dependence of the magnetic domain structures was found. Figure 5a-l (Figure S7, Supporting Information) show the AFM morphologies and corresponding MFM images of $Cr_5Te_8$ nanosheets at 2K after ZFC. We found that the thickness variation of maze-like magnetic domains. To quantitatively analyze the magnetic signal, we compared the domain width and domain contrast of these maze-like domains and found a strong thickness dependence. With increasing nanosheet thickness from 20 nm to 150 nm, magnetic domain width decreases from 850 nm to 220 nm which presents a very wide adjustable range. The corresponding variation trend is summarized in Figure 5m.

The dipole interaction dominates the exchange interaction and magnetic anisotropy in $Cr_5Te_8$, as indicated by the formation of maze-like domains.[53] Assuming ferromagnetic coupling between the $Cr_5Te_8$ layers along the $c$-axis, the nanosheets can be described by the following Hamiltonian:[54,55]

$$H = \frac{J}{2}\int [\nabla \cdot \vec{n}(\vec{x})]^2 d^2\vec{x} - K\int n_z^2(\vec{x})d^2\vec{x}$$
$$-\frac{\Omega}{4}\int \frac{[\vec{n}(\vec{x})-\vec{n}(\vec{x}')]^2 - 3\{\vec{v}\cdot[\vec{n}(\vec{x})-\vec{n}(\vec{x}')]\}}{|\vec{x}-\vec{x}'|^3}d^2\vec{x}d^2\vec{x}'$$

(1)

where $\vec{n}(\vec{x})$ is the unit vector of the magnetization at position $\vec{x}$ on a 2D plane, $J$ is the Heisenberg exchange interaction, $K$ is the overall magnetic anisotropy, $\Omega=\mu^2/a^3$ is the dipolar interaction strength ($\mu$ is the magnetic moment per spin, and $a$ is the lattice constant), and $\vec{v}=(\vec{x}-\vec{x}')/|\vec{x}-\vec{x}'|$ is the unit vector from $\vec{x}$ to $\vec{x}'$ on the 2D plane. As the dipole interaction dominating over the magnetic anisotropy, a maze-domain phase stabilized by the dipole interaction is generated, with a domain width of $\sim J/\Omega$ (on the order of $10^2$ nm for most ferromagnetic materials).[54,56] While with increasing the magnetic anisotropy and dominating over the dipole interaction, the domain width increases exponentially to eventually exceed the nanosheet size,[57] generating a single-domain phase (domain width greater than the nanosheet size) stabilized by the magnetic anisotropy. Experimentally, the rapid increase in domain width below ~40 nm suggests a transition from a maze-domain phase in the thick limit, where dipolar interaction dominates, to a single-domain phase in the ultrathin limit, where magnetic anisotropy dominates, although no single-domain nanosheets were observed.[55] This phenomenon is actually a reflection of the exchange interaction, the magnetic anisotropy and the dipolar interaction per unit area in 2D $Cr_5Te_8$, which scale differently with the nanosheet thickness.[55,58]

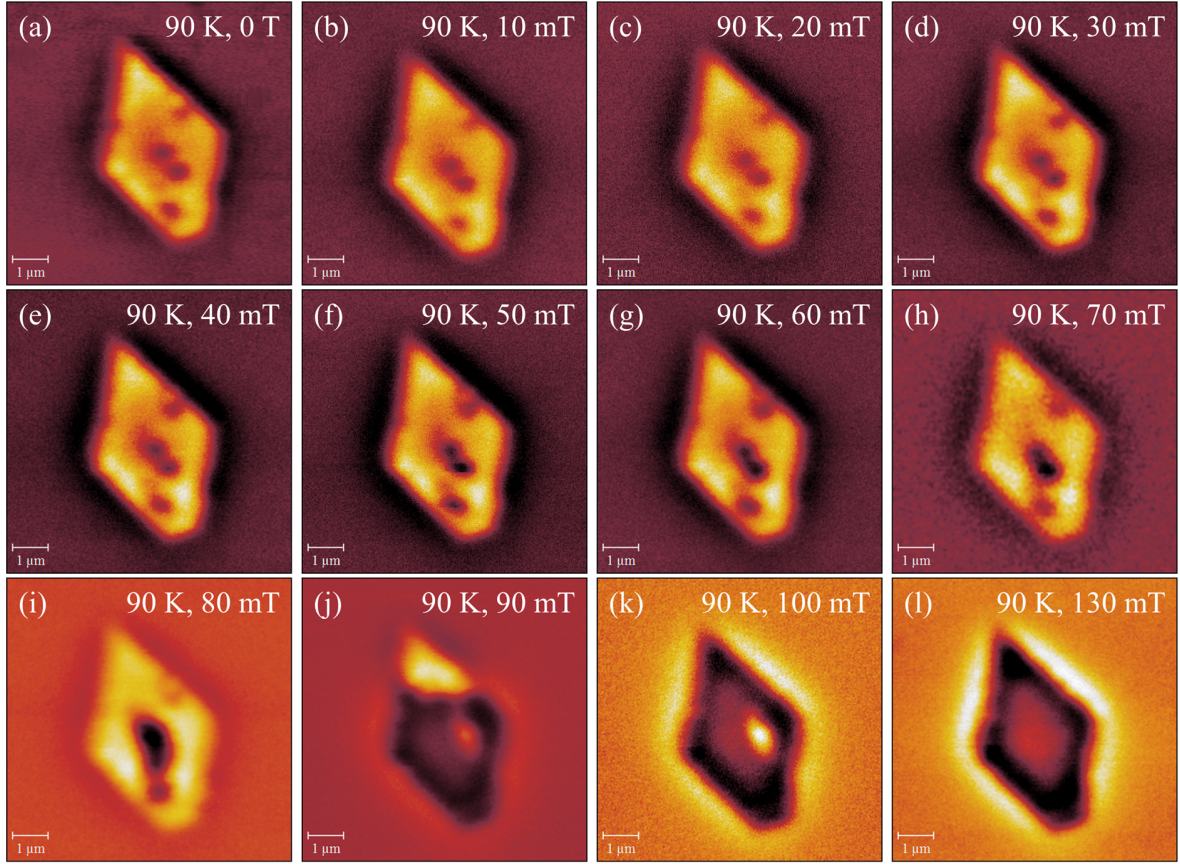

**Figure 6.** Magnetic bubbles and evolution of 2D $Cr_5Te_8$. (a) Magnetic bubbles of a $Cr_5Te_8$ single nanosheet at 90 K after ZFC. (b-l) Existence stabilized (b-f), merging (g-i), expansion (j), inversion (k) and saturation (l) of magnetic bubbles with the external magnetic field ranged from 0 mT to 130 mT. The color scales are 1.4 (a, h, k, l), 1.1 (b, c), 1.2 (d, e, f), 1.3 (g), 3 (i) and 2.8 (j) deg.

The domain contrast increases significantly with increasing the nanosheet thickness and finally becomes stable while the thickness exceeds 90 nm. When the thickness is 20 nm, the domain contrast is only 1.4 deg, and when the thickness is 90 nm, the domain contrast rapidly increases to 19.0 deg. While the nanosheet thickness exceeds 90 nm, due to the limited probe depth of the MFM tip, the contribution of the Cr atom at the bottom of the nanosheet to the magnetic signal becomes too weak. As a result, the domain contrast tends to be stable with increasing thickness which is larger than 90 nm. The corresponding thickness-dependent MFM signals is summarized in Figure 5n. According to what we know, this is first time to report the strong thickness-dependent behavior of the domain width and domain contrast, which provides a good prospect for the practical application of 2D $Cr_5Te_8$ in magnetic memory devices and other fields, and also provides a reference for the study of the magnetic domain structure of other chromium-based tellurides.

In addition to the magnetic domain structure of 2D $Cr_5Te_8$ at 2 K, we also studied the magnetic domain structure at 90 K by MFM and found another magnetic structure of 2D $Cr_5Te_8$, magnetic bubbles. We characterized the magnetic domain structure of a single nanosheet with a thickness of 40 nm by MFM after ZFC, whose corresponding AFM morphology is present in Figure 5e. Figure 6a-l (Figure S8, Supporting Information) show the MFM images of the studied $Cr_5Te_8$ nanosheet under different additional magnetic fields. We found that the magnetic domain structure of $Cr_5Te_8$ nanosheet cooled to 90 K from room temperature after ZFC was completely different from that cooled to 2K, as shown in Figure 5f and 6a. The magnetic domain structure at 2 K was a maze-like domain, while the domain structure at 90 K shows that several small magnetic bubbles with dark contrast appear within the large domain with bright contrast, and the diameter of each magnetic bubble is approximately 500 nm, which may be skyrmions. It was also discovered recently in other chromium-based tellurides such as $Cr_{1+\delta}Te_2$ (δ≈0.3),[59] $Cr_3Te_4$[60] and $Cr_{0.87}Te$.[61] Then we observed the evolution of the magnetic bubbles with an external magnetic field. With an upward additional magnetic field vertically to the *ab* plane applied to the nanosheet, we found that when the value of the additional magnetic field was within the range of 0 T to 70 mT, the magnetic bubbles existed stably. Based on their stability, we think that the magnetic bubbles are probably skyrmions.[59-61] As the additional magnetic field increased to 80 mT, the two magnetic bubbles in the center merge together. When the additional magnetic field is increased to 90 mT, the dark domain expands to cover most of the region of the $Cr_5Te_8$ nanosheet, and a new inverse magnetic bubble is formed in the dark domain. When the additional magnetic field is increased to 100 mT, the expanded dark domain completely covers the entire $Cr_5Te_8$ nanosheet and the reverse magnetic bubble is stable in the additional magnetic field of 90 mT to 120 mT. Until the additional magnetic field continues to increase to 130 mT, the reverse magnetic bubble disappears, and the magnetic domain of the $Cr_5Te_8$ nanosheet reaches saturation, forming a single domain. Our results provide the first observation of the structure and evolution of magnetic bubbles in 2D $Cr_5Te_8$, which broadens the large family of magnetic bubbles materials and provides an important reference for studying other chromium-based tellurides.

## 3. CONCLUSIONS

In summary, we have successfully synthesized high quality $Cr_5Te_8$ nanosheets and observed two magnetic domain structures: maze-like magnetic domains and magnetic bubbles. Simultaneously, the growth temperature and source–substrate distance are the key growth parameters to modulate the size of $Cr_5Te_8$ nanosheets. The $Cr_5Te_8$ nanosheets exhibit a lateral size of up to ≈300 μm and a thickness as low as ≈1.2 nm. Furthermore, the synthesized $Cr_5Te_8$ nanosheets possess both out-of-plane and in-plane ferromagnetism with Curie temperature of 179 K, which is lower than those previously reported in bulk samples. Further MFM characterizations successfully observed thickness-dependent magnetic behavior of the maze-like magnetic domains and evolution of magnetic bubbles in $Cr_5Te_8$ nanosheets. The maze-like magnetic domains originate from alternating up and down magnetic domains, whose width increasing rapidly with decreasing sample thickness. It indicated dipolar interaction played important role. We believe this work paves the way for controllable growth of 2D magnetic materials by CVD and may provide a new platform for 2D-limited magnetic domain structures and practical applications in magnetic memory devices.

## 4. EXPERIMENTAL METHODS

**Growth of 2D $Cr_5Te_8$ nanosheets.** The growth of 2D $Cr_5Te_8$ nanosheets were carried out in a two-zone tube furnace equipped with a 1-inch diameter quartz tube by an atmospheric pressure chemical vapor deposition (APCVD). Tellurium powder (400 mg, Alfa Aesar, purity 99.99%) was placed upstream in the first zone which was heated to 550 °C. And the mixed powders of $CrCl_3$ (8 mg, Alfa Aesar, purity 99.9%) and NaCl (1 mg, Alfa Aesar, purity 99.99%) were placed in the second zone, the growth temperature was set within the range of 650-850 °C. Then a freshly cleaved mica substrate (fluorophlogopite ($[KMg_3(AlSi_3O_{10})F_2]$)), a fresh and clean $SiO_2$/Si substrate (280-nm thick $SiO_2$) or a sapphire substrate, used as growth substrate, was kept near the powder of $CrCl_3$. Prior to growth, the quartz tube was vacuumed and purged by Ar gas twice to remove the residue of oxygen and moisture. A 95 sccm argon and 5 sccm hydrogen gas, respectively, was used to transport the vapor species to the downstream substrates and the growth time was 10 min under ambient pressure.

**Transfer of 2D $Cr_5Te_8$ nanosheets.** The as-grown $Cr_5Te_8$ nanosheets on mica were transferred to the target substrates ($SiO_2$/Si or gold-plated $SiO_2$/Si) using polystyrene (PS) as a medium for further characterization and MFM measurement. Briefly, $Cr_5Te_8$ nanosheets grown on mica substrates were first spin-coated with PS solution at a speed of 3000 rpm for 1 min, then baked on a hot plate at 60 °C for 30 min to improve the adhesion between the nanosheets and PS.

The edges of the PS film were then scraped off with tweezers before being placed on the water surface. After the PS/Cr$_5$Te$_8$ was successfully transferred to the target substrate, it was baked at 80 °C for 20 min. Finally, the PS was removed with acetone.

**Structure and Composition Characterization.** The morphology and thickness of Cr$_5$Te$_8$ nanosheets were characterized by OM (6XB-PC, Shang Guang), SEM (NOVA NANOSEM450, FEI) and AFM (Dimension ICON, Bruker). The phase structure of the obtained nanosheets was analyzed by XRD (D8 ADVANCE, Bruker). The elemental composition and distribution were evaluated by EDS (X-MaxN 50 mm$^2$, Oxford Instruments) equipped with SEM. The vibration modes were ascertained with a confocal Raman microscopy (alpha300 R, WITec). The elemental composition of Cr$_5$Te$_8$ nanosheets was analyzed by XPS (ESCALAB 250Xi, ThermoFisher Scientific).

**Magnetic Property Measurements.** The measurements of magnetic susceptibility and Curie temperature ($T_C$) were performed in a magnetic property measurement system (MPMS3, Quantum Design), in which the anisotropic magnetic properties of samples grown on mica substrates were observed separately. The temperature-dependent magnetic susceptibility for out-of-plane and in-plane magnetic fields was measured within the temperature range from 1.8 K to 300 K by the processes of zero-field cooling and field cooling with a field of 1000 Oe respectively. And the field-dependent magnetization studies were carried out with applied field range from −20000 Oe to +20000 Oe at temperatures of 2 K, 100 K and 200 K for out-of-plane and in-plane in several. Besides, the measurement was performed in DC mode, in which the samples were scanned vertically 30 mm in 4 s, and the gradients of magnetic field and temperature were set as 100 Oe s$^{−1}$ and 2 K min$^{−1}$ with the same intervals of 100 Oe and 0.2 K, respectively.

**Magnetic Force Microscopy Measurements.** The MFM experiments were captured by a commercial magnetic force microscope (attoAFM I, attocube) using a commercial magnetic tip (Nanosensors, PPP-MFMR, Quality factor around 1800 at 2 K) based on a closed-cycle He cryostat (attoDRY2100, attocube). The scanning probe system was operated at the resonance frequency of the magnetic tip, approximately 75 kHz. The MFM images were taken in constant height mode with the scanning plane nominally ~150 (200) nm above the substrate surface. The MFM signal, i.e., the change in the cantilever phase, is proportional to the out-of-plane stray field gradient. The dark (bright) regions in the MFM images represent attractive (repulsive) magnetization, where the magnetization is parallel (antiparallel) to the magnetic tip moments.

## AUTHOR CONTRIBUTIONS

H. X. W. and J. F. G. contributed equally to this work.

## NOTES

The authors declare no competing financial interest.

The data sets generated and/or analyzed during this study are available from the corresponding author upon reasonable request.

## ACKNOWLEDGMENT

This project is supported by the National Natural Science Foundation of China (NSFC) (No. 61674045), the Ministry of Science and Technology (MOST) of China (No. 2016YFA0200700), the Strategic Priority Research Program and Key Research Program of Frontier Sciences (Chinese Academy of Sciences, CAS) (No. XDB30000000, No. QYZDB-SSW-SYS031), and the Fundamental Research Funds for the Central Universities and the Research Funds of Renmin University of China (No. 21XNLG27 and 22XNH097).

**TOC GRAPHIC**

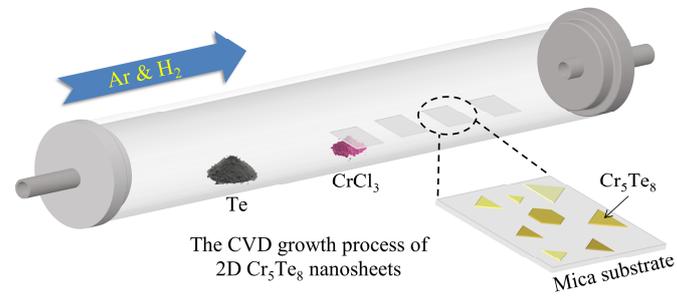

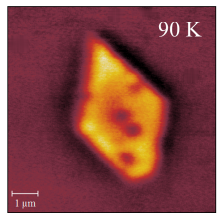
Magnetic Bubbles

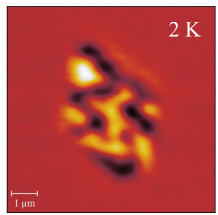
Maze-Like Domains

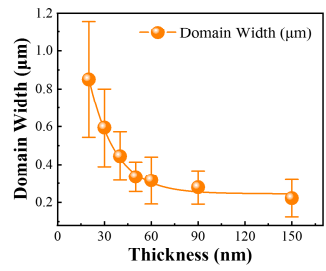

33